\documentclass[runningheads]{llncs}
\usepackage{graphicx}
\usepackage{array}
\usepackage{verbatim}
\usepackage{multirow}
\usepackage{multicol}
\usepackage{tablefootnote}
 \usepackage{hyperref}

\usepackage{amsmath,amsthm,mathtools}
\DeclareMathOperator*{\argmax}{arg\ max}

\newcolumntype{L}[1]{>{\raggedright\let\newline\\\arraybackslash\hspace{0pt}}m{#1}}
\newcolumntype{C}[1]{>{\centering\let\newline\\\arraybackslash\hspace{0pt}}m{#1}}
\newcolumntype{R}[1]{>{\raggedleft\let\newline\\\arraybackslash\hspace{0pt}}m{#1}}

\begin{document}
\title{Domain Generalization with Adversarial Intensity Attack for Medical Image Segmentation}
\author{ID: 2422}
\titlerunning{AdverIN}
%
\author{Zheyuan Zhang\inst{1} \and
Bin Wang\inst{1} \and
Lanhong Yao\inst{1} \and
Ugur Demir\inst{1} \and
Debesh Jha\inst{1} \and 
Ismail B. Turkbey\inst{2} \and
Boqing Gong\inst{3} \and
Ulas Bagci\inst{1,*}}
%
%
\institute{}
\institute{ Northwestern University, Chicago IL 60611, USA\and
National Institutes of Health, Bethesda MD 20892, USA\and
Google Research, Seattle WA 98109, USA\\
\email{ulas.bagci@northwestern.edu}}
\maketitle 
\begin{abstract}
Most statistical learning algorithms rely on an over-simplified assumption, that is, the train and test data are independent and identically distributed. In real-world scenarios, however, it is common for models to encounter data from new and different domains to which they were not exposed to during training. This is often the case in medical imaging applications due to differences in acquisition devices, imaging protocols, and patient characteristics. To address this problem, domain generalization (DG) is a promising direction as it enables models to handle data from previously unseen domains by learning domain-invariant features robust to variations across different domains. To this end, we introduce a novel DG method called Adversarial Intensity Attack (\textit{AdverIN}),  which leverages adversarial training to generate training data with an infinite number of styles and increase data diversity while preserving essential content information. We conduct extensive evaluation experiments on various multi-domain segmentation datasets, including 2D retinal fundus optic disc/cup and 3D prostate MRI. Our results demonstrate that \textit{AdverIN} significantly improves the generalization ability of the segmentation models, achieving significant improvement on these challenging datasets. Code is available upon publication.

\end{abstract}
\section{Introduction}
Domain shift, or distribution shift, refers to a situation where the statistical properties of the data change between the training and testing phases of a machine learning model. Most deep learning methods suffer from domain shift problems because the source domain for training and target domain for testing do not share the same distribution \cite{wang2022survey,zhou2022survey}. For instance, different acquisitions, scanning protocols, and patient characteristics can degrade performance of medical image segmentation tasks. A straightforward solution to address domain shift problem is to collect large-scale data from various data centers. By training the machine learning model on such diverse and extensive datasets, one can extract domain-invariant features and create a more robust model against domain shift, leading to a more generalized model. However, collecting a large-scale dataset with extensive and costly labeling is infeasible for medical fields.

Another effective method to address the domain shift is collecting some data without labeling from the target domain to adapt a source-domain-trained model. This strategy is called \textit{domain adaptation} (DA) and received considerable attention in the literature \cite{zhou2022survey}; however, target data is unavailable in advance due to privacy or even unknown before deploying the machine learning model. Therefore, DA is not a practical solution for achieving generalizable and robust performance in medical tasks.
\textbf{\textit{Domain Generalization} (DG)} is introduced to overcome the domain shift problem with absence of target data. The objective of DG is to learn a model using data from single or multiple related but distinct source domains in a way that the model can generalize well to any test domain. Current DG methods can be broadly classified into three categories: data or feature manipulation, representation learning, and various learning strategies.

Data manipulation methods involve augmenting the training data or using adversarial training to enhance the generalization capability of machine learning models \cite{liu2021feddg,xu2021randconv,zhang2020bigaug}. For example, BigAug~\cite{zhang2020bigaug} demonstrates that stacking multiple data augmentation techniques can significantly improve a model's generalization ability. Adversarial training methods can help explore more potential data space, which can help remove domain information implicitly \cite{wang2021gan}. Feature manipulation methods aim to share the style information across different domains or explore more potential feature space during training, like MixStyle, DSU~\cite{chen2022maxstyle,li2022dsu,nuriel2021pAdaIN,zhou2021mixstyle,zhang2022csu}. Representation learning methods, in contrast, explicitly force models to learn domain-invariant features through cross-domain feature alignment or disentangling features into domain-shared and domain-specific parts, such as AlignMMD~\cite{Li_2018_alignmmd} and AlignCE~\cite{morerio2018alignce}. Learning strategies, such as meta-learning and gradient operation, can also provide non-trivial improvements in a model's generalization ability. Examples of these methods include MetaReg~\cite{liu2020meta}, RSC~\cite{huang2020rsc}, Sharpness-aware Minimization~\cite{foret2021sharpnessaware,zhuang2022surrogate}.


Although DG has attracted a major attention in computer vision tasks \cite{wang2022survey,zhou2022survey}, DG studies in biomedical image segmentation tasks remain limited. This is understandable because medical imaging has unique challenges than classical computer vision tasks. For instance, the lack of large-scale medical imaging data limits the diversity of the data and makes it difficult to generalize to new domains that were not included in the training set. Another reason is that variability in the data is significantly higher than in other fields. In MaxStyle \cite{chen2022maxstyle}, authors show a style augmentation on the feature level through the adversarial attack. In AdverBias~\cite{chen2020adverbias}, authors propose an adversarial attack method to mimic various field biases for MRI images. In ShapeMeta~\cite{liu2020meta}, authors present a meta-learning strategy to utilize a shape regularization to force the model to focus on more generalized features. While these methods are promising, capturing large domain shifts in medical imaging still remains challenging. 

\begin{figure}[!b]
 \centering
 \includegraphics[width=\textwidth]{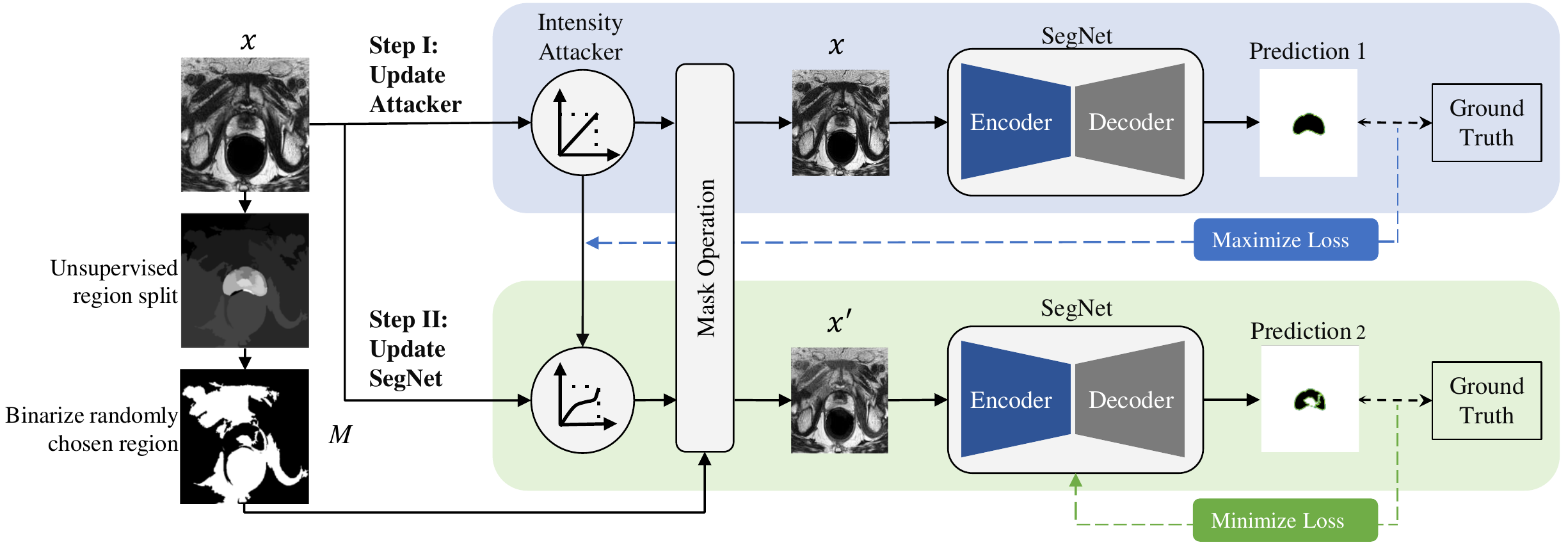}
 \caption{The proposed \textit{AdverIN} method includes two steps: first, we maximize the segmentation loss by updating the adversarial attacker's parameters, and second, we minimize the segmentation loss by updating the segmentation network's parameters.}
 \label{fig: structure}
 \end{figure}

\textbf{Summary of our contributions:} (i) We propose a novel DG method, \textit{AdverIN}, which involves an adversarial attack on the data intensity distribution to increase the model's generalization ability for the first time. Through the adversarial attack, shown in Fig.\ref{fig: structure}, we can generate the most toxic style information while preserving meaningful content information. (ii) We carefully design one monotonic intensity mapping function with mask operation, which can produce diverse data while preserving the original intensity order in local region. (iii) We demonstrate the effectiveness of our method on multi-domain 2D retinal fundus optic disc/cup and 3D prostate MRI segmentation datasets through extensive experiments and achieve significant improvement in model generalization ability. (iv) Besides, we create a large number of baseline methods for comparison, building one of the largest benchmarks for the DG in medical segmentation field. 

\section{Methods}
\vspace{-5mm}
We formulate the DG problem as follows. Let $x \in X $ represent input medical images, and $y \in Y $ be the corresponding segmentation masks. Assume $\Lambda_{s}=\{ (X, Y)^i | i=1, 2,...,D \}$ is a source domain consisting of $D$ distinct sub-domains, and the target domain $\Lambda_{t} = \{ (X, Y)^i | i=D+1, ... \}$ is inaccessible, as usual in DG problems. Let $(X, Y)^i=\{x_j^i, y_j^i\}_j^{m_i}$ represent the $i^{th}$ training domain with $j$ indexing the samples ($j=1:m_i$). Our key idea for DG is to build an intensity mapping function $x'=f(x;\rho)$ with trainable parameters $\rho$ such that it can mimic any potential intensity distribution and does not modify content in $x$.
\subsection{Adversarial Intensity Attack (AdverIN)}
We propose an adversarial training-based intensity attack method for diversifying input data. The inspiration behind this idea comes from the failure examples of medical imaging applications, where contrast and brightness can change easily in different conditions. For instance, in MRI, bias-field, non-standardness, and noise can alter the intensity distribution wildly, leading to suboptimal performance in many tasks. Thus, heavy pre-processing steps are often used to handle such changes. Herein, we design an adversarial attack strategy to mitigate the model's dependence on the absolute intensity distribution. In medical images, this intensity attack can not be arbitrary, and one must preserve local content information. Accordingly, we design an intensity attack strategy such that no matter how the intensity changes, we maintain the order of intensity, thus preserving the existence of local relative intensity information. 

\textbf{Intensity mapping function} With $\rho$ representing n+1 trainable parameters, we design the intensity mapping points as $(i/n,\hat{f}(i/n, \rho))$.  These intensity mapping points will determine the intensity adjusting function as shown in Eq.~(\ref{equation: 1}). Intensity values in this adjusting function other than discrete mapping points will be calculated through linear interpolation. Based on this function, the intensity value of every pixel in image $x$ can be adjusted according to Eq.~(\ref{equation: 2}).
 
\begin{equation}
\hat{f}(i/n; \rho) = \frac{\sum_0^i e^{\rho_j-\rho_0} - 1}{\sum_0^n e^{\rho_j-\rho_0} - 1} \in [0, 1], 
\label{equation: 1}
\end{equation}

\begin{equation}
f(x; \rho) = (x_{max} - x_{min}) \hat{f}(\frac{x - x_{min}}{x_{max} - x_{min}}; \rho) + x_{min}.
\label{equation: 2}
\end{equation}
This design strategy has two desirable properties, regardless of how the trainable parameter $\rho$ change: (i) the intensity adjusting function is monotonically increasing and (ii) it is self-bounded to original value range\footnote{The corresponding proof is provided in supplementary materials}.

\textbf{Adversarial attack} Given a segmentation network $S(x, \theta)$ and the ground truth segmentation mask  $y$, where $\theta$ represents the network parameters, every training step of \textit{AdverIN} involves two consecutive steps. Firstly, we estimate the value of intensity mapping points $\rho_0,\cdots,\rho_n$ to maximize the segmentation loss $L$ by calculating the direction of the gradient (Eq.~(\ref{equation: 3})). Secondly, we apply the intensity attacker on the input image using calculated intensity mapping points and minimize the segmentation loss on the segmentation network (Eq.~(\ref{equation: 4})). 

\begin{figure}[t]
 \centering
 \includegraphics[width=0.9\textwidth]{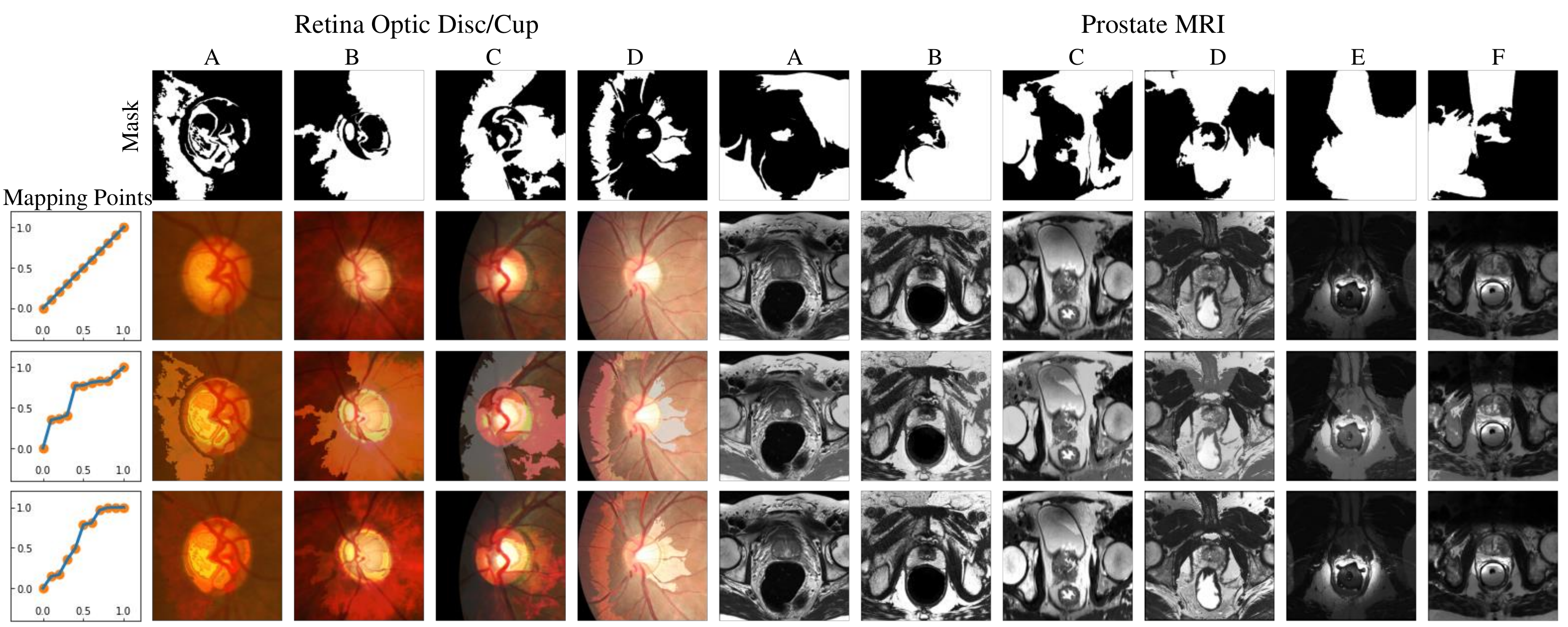}
 \caption{Visual examples of intensity attacks on optic fundus and prostate MRI datasets. Left column shows intensity mapping points and adjusting function after interpolation}
 \label{fig: influence}
 \vspace{-5mm}
 \end{figure}

Furthermore, we limit the intensity attack in the binary region mask $\mathcal{M}$, which can be generated by randomly choosing regions in unsupervised image segmentation masks\footnote{More details about $\mathcal{M}$ generation are provided in the supplementary materials}. The mask operation on the input image is expressed as $x' = \mathcal{M} \cdot f(x, \rho)+(1-\mathcal{M})\cdot x$ and when $\rho$ is initialized with 0 in the first step $x' = x$. This mask operation allows us to conduct adversarial attacks within the local regions. Fig. \ref{fig: influence} provides some intuitive examples of intensity attackers.
\begin{equation}
\hat{\rho} = \displaystyle{\argmax_{\|\rho\|_2 < \delta}} \: L( S( \mathcal{M} \cdot f(x, \rho)+(1-\mathcal{M})\cdot x, \theta), y) 
\approx \frac{\frac{\partial L}{\partial \rho}}{||\frac{\partial L}{\partial \rho}||} \delta,
\label{equation: 3}
\end{equation}
\begin{equation}
 Obj:=\displaystyle{\min_{\theta}} \: L( S(\mathcal{M}  \cdot f(x, \hat{\rho})+(1-\mathcal{M})\cdot x, \theta) , y).
\label{equation: 4}
\end{equation}

\textbf{Training details} We adopt the standard 2D ResUNet network implementation of MONAI based on PyTorch as the segmentation backbone while our approach is also applicable to other networks \cite{cardoso2022monai}. The loss function is based on the classic combination of Dice and Cross-Entropy Loss. For all experiments, we select SGD optimizer and cosine learning rate decay due to its strong domain generalization ability compared with the adaptive method, like Adam~\cite{amir2021sgd,wu2020noisy,zhou2020sgd}. We perform a relatively long training with 1000 epochs for each dataset. All experiments are conducted on 6 NVIDIA RTX A6000 GPUs.
\section{Experiments and Results}
\subsection{Datasets Preparation}
\textbf{2D Retinal Fundus}:  This dataset contains 2D Retinal Fundus images from 4 different
clinical centers (A, B, C, D) for optic disc/cup segmentation \cite{liu2020meta}. Following previous research, we center-crop an 800$\times$800 disc region and resize the cropped region to 384$\times$384. We apply Min-Max Normalization to normalize each image between -1 and 1. All preprocessing codes, organized datasets, and associated training logs are all released.

\noindent\textbf{3D Prostate Segmentation}: This dataset comprises 3D prostate MRI images obtained from various medical centers, including Nijmegen Medical Centre (A), Boston Medical Center (B), Universitat de Girona (C), University College London (D), Beth Israel Deaconess Medical Center (E), and Haukeland University Hospital (F) \cite{liu2021feddg}. We resize the images to 384$\times$384 in the axial plane and extract 2D slices with a depth of 3 from the 3D volumes. Next, we clip the data between the 0.5-99.5 percentiles and normalize each slice between -1 and 1 using Min-Max Normalization. It is worth noting that we evaluate the final performance at the case level, which includes all positive and negative slices. 

\subsection{Comparison Results}
We conduct the training and evaluation under the standard Leave-One-Domain-Out protocol \cite{zhou2022survey}. For example, given N domains in total, we use N-1 domains for training and the last unseen domain for evaluation. Thus, we can build one strong segmentation generalization baseline for comparison by combining various data from different domains. Finally, the average performance across N domains will be used for evaluating the model's generalization ability. We establish the upper bound with intradomain training for domain generalization comparison by cross-validation on all domain data. We compare our model's performance with other strong DG methods, including several categories: data augmentation: BigAug \cite{zhang2020bigaug}, RandConv \cite{xu2021randconv}, MixUp \cite{zhang2018mixup};  style augmentation: MixStyle \cite{zhou2021mixstyle}, pAdaIN \cite{nuriel2021pAdaIN}, DSU \cite{li2022dsu}. Particularly, we also evaluate the model's performance when removing style statistics in BatchNorm with InstanceNorm to compare style augmentation\cite{huang2017arbitrarystyle}; domain alignment: CrossEntropy Alignment (AlignCE) \cite{morerio2018alignce}; gradient operation: Representation Self-Challenge (RSC) \cite{huang2020rsc}. 

\begin{table}[t]
\centering
\caption{Quantitative comparison for fundus optic disc/cup segmentation task. \textit{AdverIN} significantly improves overall out-of-domain segmentation performance compared with the baseline methods and outperforms other domain generalization methods. Higher Dice(\%) and lower HD95 indicate goodness of the segmentation. Since fundus images are in png format, HD95 here is calculated in pixels}
\resizebox{0.85\columnwidth}{!}{%
\begin{tabular}{c | *{4}{ C{1.1cm}@{}} |*{4}{ C{1.1cm}@{}} |c}
  \hline
  \hline
  Task & \multicolumn{4}{c|}{Optic Disc Segmentation} &\multicolumn{4}{c|}{Optic Cup Segmentation}  &\multirow{2}{*}{Overall} \\
  \cline{0-8}
  Unseen Domain& A & B & C & D & A &B &C &D &  \\
  \hline
  \hline
  &\multicolumn{9}{c}{\textbf{Dice Coefficient (Dice)~$\uparrow$}} \\ 
  \hline
  Intradomain  & 85.87 & 85.57 & 89.84 & 91.41 &87.57 &80.89 & 87.1 & 89.65 & 87.24 \\
  \hline
  Baseline  &  80.87 &  69.76 &  83.44 &  76.61 &  85.7 & 66.49 & 82.76 & 74.11 &77.47 \\
  BigAug \cite{zhang2020bigaug}&  81.06 &  69.36 &  84.14 &  80.58 & 86.49 & 66.85 & 82.65 & 78.46 & 78.7 \\
  MixUp \cite{zhang2018mixup}&  \textbf{82.23} &68.5 &  83.34 &  77.31 & 86.63 & 67.15 & 82.61 & 75.75 &77.94 \\
  RandConv \cite{xu2021randconv}&  81.23 &  60.15 &  83.96 &  86.61 & 81.25 & 60.28 & 83.31 & 84.53 &  77.66 \\
  InstanceNorm  &  81.68 &  69.53 &  \textbf{85.45} &  81.89 & 82.81 & 66.61 & 80.79 & 79.86 &78.58 \\
  MixStyle \cite{zhou2021mixstyle}&  79.92 &  69.41 &  82.78 &  77.56 & 85.56 & 66.98 & 82.44 & 75.98 &77.58 \\
  pAdaIN \cite{nuriel2021pAdaIN}& 81.28 & 70.83 &  81.8 & 75.32 & 86.44 & 69.98 & 81.47 &73.34 & 77.56 \\
  DSU \cite{li2022dsu}&  81.34 &  68.93 &  82.73 &  77.08 & \textbf{86.73} & 66.55 & 82.85 & 75.11 &77.66 \\
  AlignCE \cite{morerio2018alignce}& 80.82 & 71.05 & 83.31 & 76.02 &86.59 &69.44 &82.02 &74.09 & 77.92 \\
  RSC \cite{huang2020rsc}& 82.17 &  70.63 &  84.73 &  81.23 & 86.45 & \textbf{70.78} & 82.89 & 77.7 & 79.57 \\
  \hline
  \textit{AdverIN} (Ours) &  82.21 & \textbf{72.21} & 84.62 &  \textbf{86.9} & 85.28 &69.71 & \textbf{83.79} & \textbf{84.35} & \textbf{81.13} \\
  \hline
  \hline
  &\multicolumn{9}{c}{\textbf{Hausdorff Distance 95 (HD95)~$\downarrow$} \footnotesize{(pixel)}} \\ 
  \hline
  Intradomain &  20.64 &  21.95 &  17.33 &11.4 & 24.11 & 23.07 & 18.42 &  10.8 &  18.47 \\
  \hline
  Baseline  & 24.78 & 42.09 & 24.61 & 27.31 &29.73 &45.62 &22.02 &32.63 & 31.1 \\
  BigAug \cite{zhang2020bigaug} & 24.13 & 45.78 & 23.75 &  23.4 & \textbf{27.88} &46.75 &22.29 &25.28 &29.91 \\
  MixUp \cite{zhang2018mixup} & 23.98 & 47.33 &  24.2 & 28.91 &28.52 &45.15 &22.11 &30.59 &31.35 \\
  RandConv \cite{xu2021randconv} & 29.34 & 69.15 & 23.75 & 18.68 &35.66 &63.66 &21.86 &16.56 &34.83 \\
  InstanceNorm & 28.94 & 45.98 & 23.13 & 28.26 &33.81 &43.85 &23.03 &25.17 &31.52 \\
  MixStyle \cite{zhou2021mixstyle} & 25.83 & 41.39 & 25.27 & 27.78 &30.54 &43.86 &22.47 &29.45 &30.82 \\
  pAdaIN  \cite{nuriel2021pAdaIN}& 24.51 & \textbf{39.45} & 25.51 & 30.57 & 29.13 & 41.82 & 22.87 &  33.35 & 30.9 \\
  DSU \cite{li2022dsu} & 23.98 & 44.01 & 25.14 &  28.4 &28.08 &43.87 &22.37 &31.14 &30.87 \\
  AlignCE \cite{morerio2018alignce} &  24.57 &  41.38 &  25.05 &  29.52 & 28.32 & 43.89 & 22.77 & 33.38 &  31.11 \\
  RSC \cite{huang2020rsc} & \textbf{23.89} &  42.6 & 23.41 & 22.76 &27.93 & \textbf{42.6} & 21.41 &26.69 &28.91 \\
  \hline
  \textit{AdverIN} (Ours) & 25.56 &  40.89 &  \textbf{22.85} &  \textbf{17.25} & 31.29 & 44.71 & \textbf{21.07} & \textbf{17.46} &  \textbf{27.64} \\
  \hline
  \hline
  \end{tabular}}
\label{tab:fundus_seg}
\vspace{-3mm}
\end{table}

\textbf{Evaluation on the retinal fundus dataset.} As summarized in Table \ref{tab:fundus_seg}, we  observe that \textit{AdverIN} significantly improves the segmentation performance on unseen domains for retinal fundus optic disc/cup compared with the baseline method. Particularly in Domain C, \textit{AdverIN} improves the Dice-Coefficient (Dice) significantly from 76.61 to 86.9 (+13.4\%) for the optic disc and 74.11 to 84.35 (+13.8\%)  for the optic cup. For DG studies, it is crucial to ensure that
the model can perform well across all domains and classes, thus, we examine the overall average performances rather than individual ones. We can note that \textit{AdverIN} improves the average Dice from 77.47 to 81.13 (+4.7\%)  and reduces the Hausdorff Distance 95 (HD95) from 31.1 to 27.64 across two classes in four domains, showing significant superiority over other methods.
 \begin{figure}[t]
 \centering
\includegraphics[width=\textwidth]{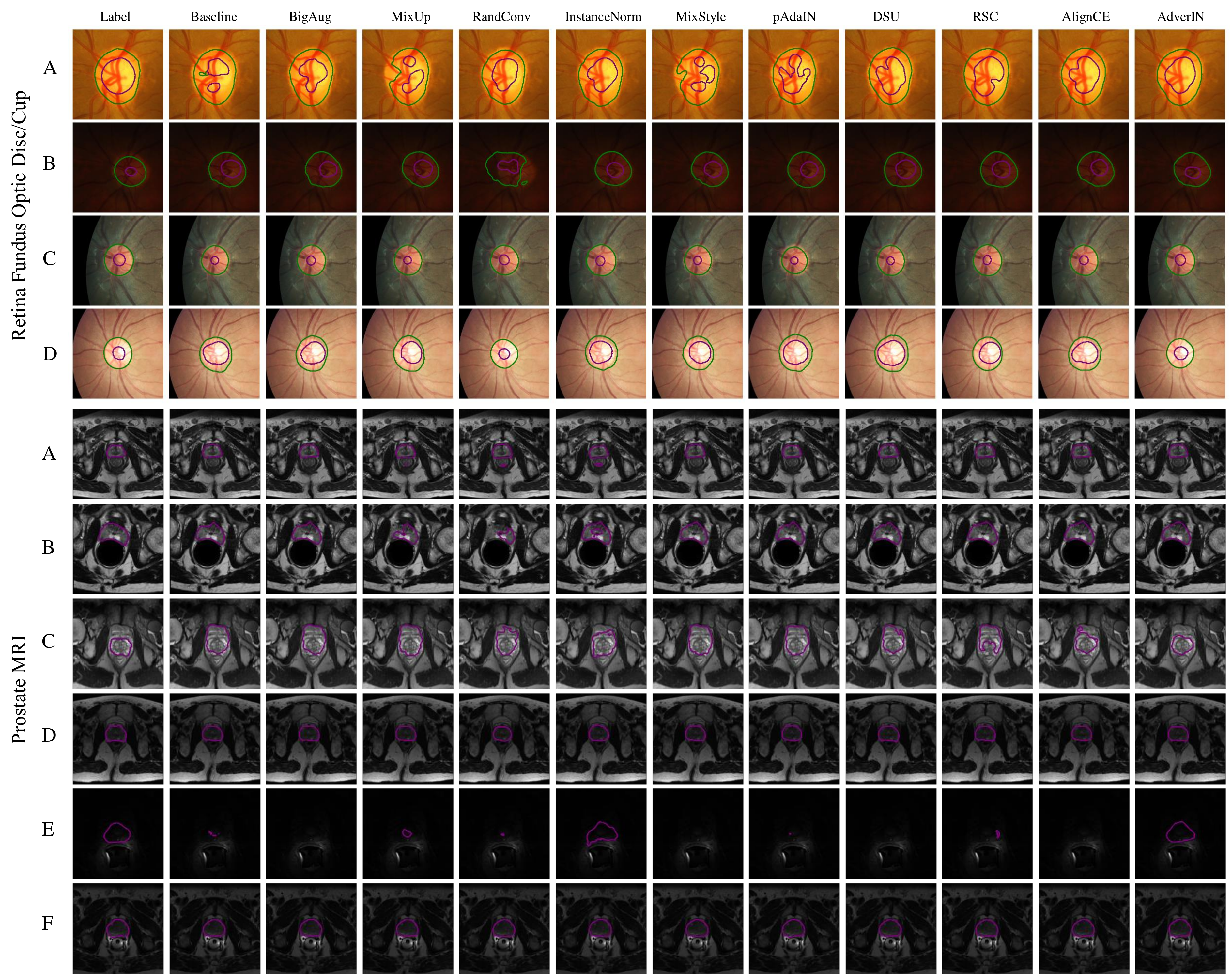}
 \caption{Visual comparison for fundus optic disc/cup and prostate MRI segmentation task. Compared with other domain generalization methods, \textit{AdverIN} can mitigate the influence of domain shift and generate more reliable segmentation.}
 \label{fig: segvisual}
 \vspace{-5mm}
 \end{figure}
 \begin{table}
    \caption{Quantity comparison for prostate MRI segmentation task. \textit{AdverIN} also significantly outperforms the baseline and other domain generalization methods. }
    \centering
    \resizebox{\columnwidth}{!}{%
    \begin{tabular}{c | *{6}{ C{1.1cm}@{}}|c| *{6}{ C{1.1cm}@{}}|c }
    \hline
    \hline
    Task & \multicolumn{14}{c}{Prostate MRI Segmentation}  \\
    \hline
    Unseen Domain  &  A &  B &  C &  D &  E &  F & \multirow{2}{*}{Overall} &  A &  B &  C &  D &  E &  F & \multirow{2}{*}{Overall}\\
    \cline{0-6}
    \cline{9-14}
    &\multicolumn{6}{c|}{\textbf{Dice Coefficient (Dice)~$\uparrow$}} &&\multicolumn{6}{c|}{\textbf{Hausdorff Distance 95(HD95)~$\downarrow$} \footnotesize{(mm)}} &\\ 
    \hline
    \hline
    Intradomain & 90.59 &  87.71 & 80.91 &87.5 & 85.43 & 87.53 &86.61 &  4.8 &  8.31 & 5.85 &  5.95 & 7.38 & 4.71 &6.17 \\
    \hline
    Baseline & 89.24 & 82.1 & 71.03 &88.59 & 61.22 &  86.63 &79.8 &5.4 & 9.91 &  12.39 & 4.52 &  12.19 &4.98 & 8.23 \\
    BigAug \cite{zhang2020bigaug} & 89.69 & 83.6 &  70.25 &88.02 &  63.57 & 86.95 & 80.35 &4.91 & \textbf{8.43} &  13.81 &  5.1 & 10.85 &4.86 &7.99 \\
    MixUp \cite{zhang2018mixup} & 87.71 &80.46 & 68.33 &85.38 & 67.04 &  86.32 &79.21 &7.28 &12.66 &  15.72 & 9.94 &  22.73 & 5.2 &12.25 \\
    RandConv \cite{xu2021randconv}&  85.7 &74.99 & 59.85 &78.51 & 55.58 &  83.78 &73.07 &7.25 &18.64 &  20.38 & 7.77 &  14.18 &5.51 &12.29 \\
    InstanceNorm  & 87.75 &81.67 & 69.56 &85.01 & 63.19 &  85.27 &78.74 &5.95 &14.23 &  15.66 &  7.6 &  25.73 &5.66 &12.47 \\
    MixStyle \cite{zhou2021mixstyle} & 88.67 &81.79 & 72.87 &88.33 & 62.67 &  85.52 &79.98 &5.3 &13.13 &  14.43 & 4.34 &12.3 &5.66 & 9.19 \\
    pAdaIN \cite{nuriel2021pAdaIN} & 88.67 &80.94 & \textbf{74.59} &87.73 & 62.65 &  86.86 &80.24 &5.45 &12.12 &  11.45 & 5.66 &  11.99 &5.18 & 8.64 \\
    DSU \cite{li2022dsu} & 89.1 &83.2 & 72.28 & 89.01 & 72.39 &  86.08 &82.01 &5.38 &12.61 &  13.79 & 4.53 &  11.48 &4.93 & 8.79 \\
    RSC \cite{huang2020rsc} & \textbf{90.74} &  \textbf{84.24} & 70.7 &  \textbf{89.03} & 67.3 & \textbf{87.03} &  81.51 & \textbf{4.72} &  9.48 &9.76 &  \textbf{4.21} &   11.11 & 5.09 &  7.39 \\
    AlignCE \cite{morerio2018alignce}  & 89.86 &83.81 & 72.27 &88.62 & 59.84 &  86.79 &80.2 &4.84 &10.84 &  10.12 & 4.45 &  11.03 & \textbf{4.86} & 7.69 \\
    \hline
    \textit{AdverIN} (Ours) & 89.62 &  83.89 &72.38 &  87.77 & \textbf{74.84} & 85.87 &  \textbf{82.4} &5.23 & 10.18 & \textbf{8.23} &  5.35 & \textbf{8.76} & 5.58 &  \textbf{7.22} \\
    \hline
    \hline
    \end{tabular}}
\label{tab:prostate_seg}
\vspace{-2mm}
\end{table}

\textbf{Evaluation on the prostate MRI dataset.} As summarized in Table \ref{tab:prostate_seg}, \textit{AdverIN} also achieves impressive advances in the segmentation performance on unseen domains for the prostate MRI dataset. Particularly in Domain E, \textit{AdverIN} improves Dice from 61.22 to 74.84 (+22.2\%)  for the prostate. Interestingly, some domain generalization algorithms like pAdaIN, and DSU achieve solid improvements in the Dice metric but degrade the performance in the HD95 metric, indicating a failure in the boundary identification while biased in the region terms. In comparison, \textit{AdverIN} improved the Dice score from 79.8 to 82.4 (+3.3\%)  and reduced HD95 from 8.23 to 7.22 across 6 domains, further proving the designed model's effectiveness. Another interesting observation is that many methods like MixStyle, pAdaIN, DSU can achieve impressive improvement in one dataset but degrade performance in another. BigAug and RSC can consistently improve results on both datasets, while AdverIN achieves stable superiority.

\subsection{Ablation Experiments} 
\noindent \textbf{Influence of Intensity Mapping Points Number}: We conduct ablation experiments to seek the influence of intensity mapping function design parameters: number of intensity mapping points. Different intensity mapping point numbers can lead to different intensity adjusting functions in Eq.\ref{equation: 1}. Our ablation results demonstrate that, regardless of the number of intensity mapping points, \textit{AdverIN} consistently improves over naive domain combinations (Dice: 77.47). Additionally, we observe that setting the intensity mapping points number too low or too high can impede achieving the best performance. 

\noindent \textbf{Influence of Adversarial Attacking Intensity}: In another ablation experiment, we investigate the impact of hyper-parameter $\delta$ which determines the intensity of the adversarial attack on the segmentation model's performance (see Eq.\ref{equation: 3}). We observe that there exists one stable plateau when setting the attack intensity between 1-5, and an out-range setting limits the performance gain.
\begin{figure}
 \vspace{-5mm}
 \centering
 \label{fig: segvisual}
 \includegraphics[width=\textwidth]{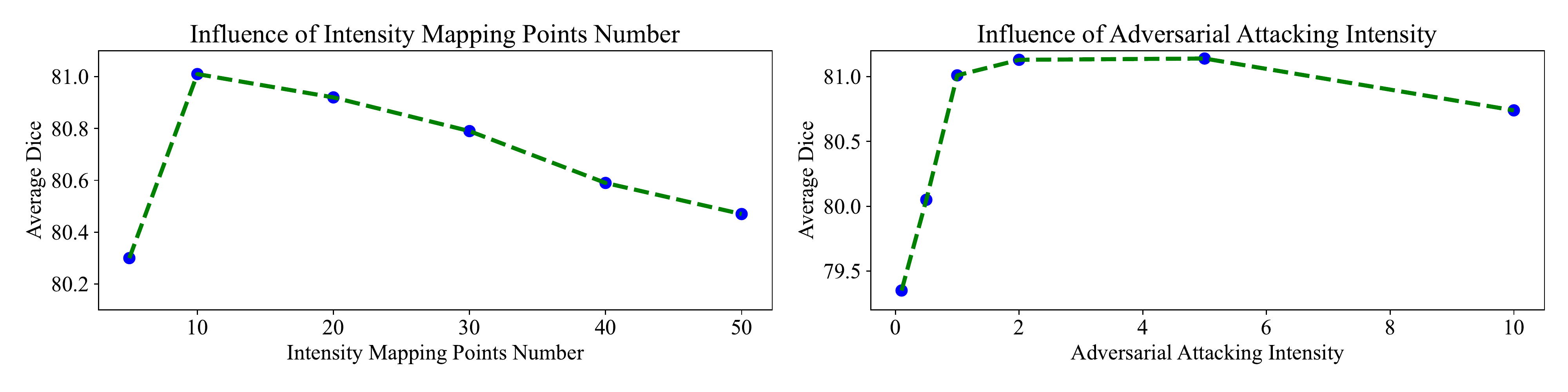}
 \caption{Ablation Experiments on 2D retina fundus optic disc/cup segmentation tasks. We separately investigate the influence of intensity mapping points number and adversarial attack intensity.}
 \label{fig: ablation}
 \vspace{-8mm}
 \end{figure}
\section{Conclusion}
In this study, we propose a novel domain generalization method for medical image segmentation via Adversarial Intensity (\textit{AdverIN}) attack by increasing data diversity. This new adversarial data augmentation method generates training samples with infinite number of styles while preserving the essential content information, which helps to remove domain-dependent style information during model training. Our proposed method offers a promising solution to the DG problem in medical fields and achieves 
significant improvements for optic disc/cup and prostate MRI segmentation from multiple domains. 

%
%
%
%
\newpage
\bibliographystyle{splncs04}
\bibliography{refer}

\end{document}


\noindent \large \textbf{Supplementary Materials}

\section{Arbitrary Mask Generalization}

One important component of AdverIN to increase data diversity is generating binary region masks, which can limit the intensity of attacking within the local region. To guarantee meaningful local region split, we adopted the image segmentation tools provided by Scikit-Image based on  k-means clustering. We cluster the whole image, including foreground and background, into 20 regions in total. During the training, we will randomly choose five from 20 regions. This region split is conducted in the data preprocessing stage, which will not increase further computation during training.  The visual split examples are provided below, and the region-splitting code is also released with the project.
\begin{figure}
 \centering
 \includegraphics[width=0.85\textwidth]{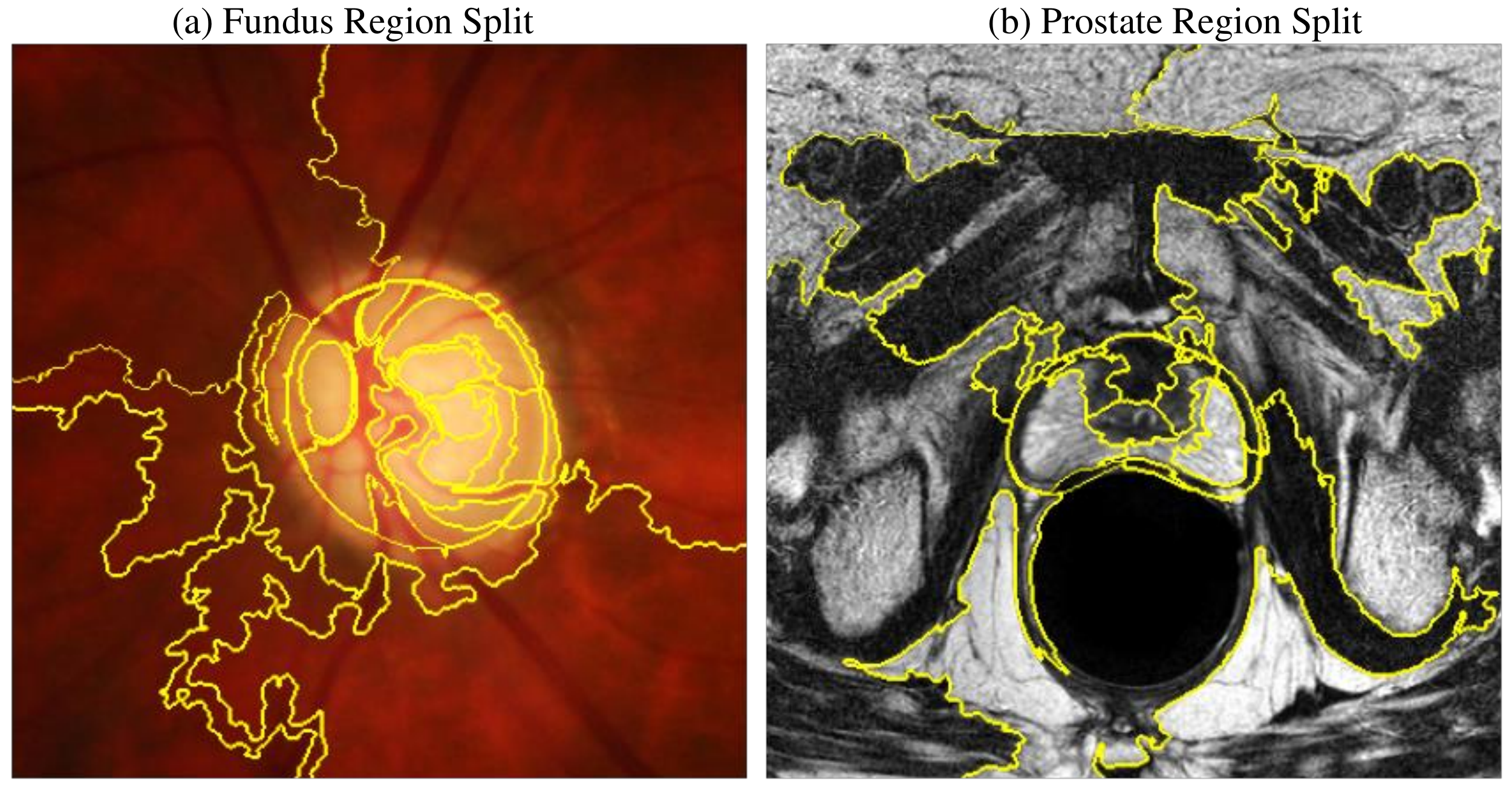}
 \caption{Region split examples for a) retina fundus optic disc/cup b) prostate MRI based on k-means clustering.}
 \label{fig: ablation}
 \end{figure}

\section{Proves for Monotonic and Self-Bounded}
We design the intensity mapping function with two good properties: the intensity adjusting function is monotonically increasing and self-bounded to the original data range. These two properties are essential for maintaining order in the local region. Here we proves that Equation 1 follows such properties:

\noindent \textbf{Monotonic} Given two inputs $i_1, i_2 \in \{0..n\} $ and $i_2>i_1$, we will have that
\begin{gather*}
     \hat{f}(i_1/n, \rho) = \frac{\sum_0^{i_1} e^{\rho_j-\rho_0} - 1}{\sum_0^n e^{\rho_j-\rho_0} - 1} \\
     \hat{f}(i_2/n, \rho) = \frac{\sum_0^{i_2} e^{\rho_j-\rho_0} - 1}{\sum_0^n e^{\rho_j-\rho_0} - 1} \\
     \hat{f}(i_2/n, \rho) - \hat{f}(i_1/n, \rho) = \frac{\sum_{i_1+1}^{i_2} e^{\rho_j-\rho_0}}{\sum_0^n e^{\rho_j-\rho_0} - 1}
\end{gather*}
Due to $i_2>i_1$, thus $i_2>=(i_1 +1 )$, we will have
\begin{gather*}
     \hat{f}(i_2/n, \rho) - \hat{f}(i_1/n, \rho) = \frac{\sum_{i_1+1}^{i_2} e^{\rho_j-\rho_0}}{\sum_0^n e^{\rho_j-\rho_0} - 1}>=0
\end{gather*}
\noindent \textbf{Self-Bounded} Given that the function is monotonically increasing
\begin{gather*}
    \ max \quad \hat{f}(i/n, \rho) = \hat{f}(n/n, \rho)\\
    \ max \quad \hat{f}(i/n, \rho) = \frac{\sum_0^{n} e^{\rho_j-\rho_0} - 1}{\sum_0^n e^{\rho_j-\rho_0} - 1}\\
    thus \quad \ max \hat{f}(i/n, \rho) = 1 \\
    \ min \quad \hat{f}(i/n, \rho) = \hat{f}(0/n, \rho)\\
    \ min \quad \hat{f}(i/n, \rho) = \frac{\sum_0^{0} e^{\rho_0-\rho_0} - 1}{\sum_0^n e^{\rho_j-\rho_0} - 1} \\
    thus \quad \ min \hat{f}(i/n, \rho) = 0
\end{gather*}
Therefore, we could get that no matter how $\rho$ varies, $\hat{f}(i/n, \rho)$ is always monotonically increasing and self-bounded to $[0, 1] $

\section{Comparing with Intra-Domain Training}

Although most domain generalization methods can advance the baseline method, we still want to show one significant gap exists compared with intra-domain training. As shown in Tables 1, 2, it is unsurprising that intra-domain can outperform all domain generalization algorithms. For the fundus optic disc/cup segmentation dataset, this gap is from 81.13 to 87.24 in Dice. For prostate MRI segmentation, this gap is from 82.40 to 86.61 in Dice. This phenomenon inspires us to propose more advanced algorithms in the future.